\documentclass[12pt,]{article}
\usepackage{amsmath}
\usepackage{amsfonts} \usepackage{amssymb}
\newtheorem{theorem}{Theorem}[section]

\newtheorem{proposition}[theorem]{Proposition}

\numberwithin{equation}{section}

\begin{document}      

\title{Smooth non-zero rest-mass evolution across time-like infinity}

\author {Helmut Friedrich\\
Max-Planck-Institut f\"ur Gravitationsphysik\\
Am M\"uhlenberg 1\\ 14476 Golm, Germany }

\maketitle                
 
\begin{abstract}
It is shown that solutions to Einstein's field equations with positive cosmological constant 
can include non-zero rest-mass fields which coexist with and travel unimpeded across a smooth conformal boundary. This is exemplified by the coupled Einstein-massive-scalar field equations for which the mass $m$ is related to the cosmological constant $\lambda$ by the relation
$3\,m^2 = 2\,\lambda$. 
Cauchy data for the conformal field equations can in this case  be prescribed 
 on the (compact, space-like) conformal boundary ${\cal J}^+$. Their developments backwards in time induce a set of standard Cauchy data on space-like slices for the Einstein-massive-scalar field equations which is open in the set of all Cauchy data for this system.

\end{abstract}

\newpage

{\footnotesize

\section{Introduction}

In his work on {\it conformal cyclic cosmology} \cite{penrose:2010} Roger Penrose conjectures the ocurrence of concentric circles in the CMB which reflect  bursts of gravitational radiation resulting from encounters of supermassive black holes in an aeon preceding the big bang of our present one.  A recent  analysis \cite{gurzadyan:penrose:2013} of the CMB based on WMAP data  and an independent, even more recent, study  of the CMB maps observed by the Planck collaboration \cite{an:meissner:nurowski:2013}
 indeed seem to identify ring like structures in the CMB sky. While these findings indicate strong support of Penrose's proposal, the theoretical reasoning which led to them still raises questions. 

The underlying picture is that of a smooth, time oriented conformal structure ${\cal C}$  of signature $(-,+,+,+)$ on a $4$-dimensional manifold ${\cal M} \sim \mathbb{R} \times S$ with  compact $3$-manifold $S$,  into which an infinite sequence of {\it aeons},  time oriented `physical'  solutions to Einstein's field equations with cosmological constant $\lambda > 0$, are conformally embedded so that any two neighbouring aeons are separated by a {\it crossover $3$-surface} ${\cal X} \sim S$ which is space-like with respect to the conformal structure.
The aeons start with a big bang that `touches' the preceding crossover surface  while their  future end represents an  exponentially expanding space-time for which the following 
crossover surface defines a smooth conformal boundary in the sense proposed by Roger Penrose in \cite{penrose:1965}. 

Consider the Einstein equations with cosmological constant $\lambda > 0$ and  vanishing energy momentum tensor
and denote by ${\cal D}_S$ the set of Cauchy data (i.e. solutions to the constraints) for these equations on $3$-manifolds diffeomorphic to $S$. Not all the space-times developing  from these data  admit  a smooth conformal boundary diffeomorphic to $S$. Notable examples are the Einstein-de Sitter solution, which includes black holes and admits only patches of a smooth conformal boundary \cite{gibbons:hawking:1977}, and the Nariai solution, which is geodesically complete but does not even admit a piece of a smooth conformal boundary \cite{beyer:2009}. 
Nevertheless, the class of solutions which do admit a smooth conformal boundary diffeomorphic to $S$ is fairly rich. This  is a consequence of a peculiar feature  
of Einstein's field equations. In the vacuum case
they admit representations in terms of the conformal fields, referred to as  the {\it conformal fields equations} (see section \ref{cfe}), which induce under suitable gauge assumptions equations that are hyperbolic even where the conformal factor vanishes or becomes negative \cite{friedrich:asymp-sym-hyp:1981b},
\cite{friedrich:1991}.

The conformal  equations can be solved backwards in time with Cauchy data which are prescribed on the future conformal boundary ${\cal J}^+ \sim S$. The freedom to prescribe data there  is essentially the same as in the standard Cauchy problem, though,  due to the fact that the conformal boundary is geometrically a very special hypersurface relative to the solution space-time, there are differences in the interpretation of the data \cite{friedrich:1986a}. 
Let ${\cal A}_S$ denote the Cauchy data pertaining to solutions obtained by such backward developments. It turns out that  ${\cal A}_S$ is an open subset of ${\cal D}_S$  (if endowed with  a natural Sobolev topology). This follows by the argument used in \cite{friedrich:1986b} to show the non-linear stability of de Sitter space.
In fact, due to their hyperbolicity, the conformal field equations see in principle no difference between backward and forward evolution. The data on the conformal boundary which have been evolved backwards can thus also be evolved forwards into domains foliated by Cauchy hypersurfaces on which the conformal factor is negative. The resulting solutions  to the conformal field equations then extend smoothly across the conformal boundary. 
Cauchy stability for hyperbolic systems then implies that data in ${\cal D}_S$ which are sufficiently close to ${\cal A}_S$ also develop into domains where the conformal factor is negative and the field equations themselves then ensure that the set where the conformal factor vanishes defines a smooth conformal boundary for the vacuum space-times arising from the given data in  ${\cal D}_S$ (which are thus seen to be in fact in ${\cal A}_S$).

The solution to the conformal field equation in the future of the conformal boundary again defines a solution to the Einstein equations we started with. This solution is also  determined uniquely by the Cauchy data we prescribed in the past of the conformal boundary.

These results generalize to matter fields whose energy momentum tensor is trace free and which obey conformally covariant field equations.
This has been exemplified in detail  in \cite{friedrich:1991}  by the Maxwell or the Yang-Mills equations but it holds true for other such equations. It follows that if a solution to these equations does admit a smooth conformal boundary in the future, this is true also for all solutions which are, in terms of Cauchy data on some given time slice, close to the given one. In all these cases gravitational radiation or other field excitations will travel unimpeded across the conformal boundary.

There remains the question of what will happen to the prospective conformal boundary in the presence of  fields with non-vanishing rest mass.  In \cite{penrose:2010} it has been assumed that there will be  some past neighbourhood  of the crossover surface in which only zero rest mass fields will be present. While it is far from obvious how massive fields behave at the end of an unlimited spatial expansion, this certainly  seems to be a strong requirement. At present no process is known which would allow one  to justify it. It is the main purpose of the following analysis  to show that this restriction may  not be necessary.

There is a second problem, which arises right at the crossover surfaces.
The solution to the Einstein equations obtained in the future of a crossover surface ${\cal X}$ by 
 extending the solution to the conformal field equations smoothly across ${\cal X}$  
will start to contract and thus rather resemble a  time reversed version of an exponentially expanding space-time instead of a big bang solution as required by the standard scenario. It is suggested therefore in \cite{penrose:2010} that in the immediate 
future of ${\cal X}$ each aeon space-time evolves instead according to the  `isotropic cosmological singularity' model studied by Paul Tod (cf. \cite{tod:2003}, which also gives references to earlier work in this direction). 
In this setting it is assumed that the space-time admits a conformal rescaling which 
blows up the space-time near the big bang so that the latter can be represented by a 
space-like hypersurface smoothly attached to the past end of the original space-time. The idea then is to identify  this hypersurface  and the fields obtained on it by the blow up procedure with the preceding crossover surface and the data induced on that by the smooth conformal extension from the previous aeon.

There appears to be a basic difference, however,  between the `blow down' procedure underlying the construction of the conformal boundary considered above and the blow up procedure defining the 
isotropic cosmological singularity picture.
The latter seems to admit no version of conformal field equations which induce hyperbolic evolution equations near the conformal boundary under fairly general assumptions. In fact, the freedom to prescribe initial data for the future evolution on the past boundary turns out to be rather restricted \cite{tod:2003}. Evolving such restricted data into the future, performing a slight generic perturbation of the data induced on some Cauchy hypersurface, and then evolving backwards  will, more likely than not, result in a space time which does not admit a conformal blow up leading to a smooth setting.

In general it is not clear to what extent data induced on ${\cal X}$ from the previous aeon can be evolved further  in the new setting and if they can the extension procedure will not be stable. 
Moreover, it is not clear which mechanism should convert on ${\cal  X}$ the evolution law carried across ${\cal X}$  with the conformal field equations  into an evolution law consistent with the isotropic cosmological singularity setting. 
This problem will be addressed again in section \ref{hom-sols}.

\vspace{.2cm}

Required is  an insight into the asymptotic behaviour at time-like infinity of fields with non-vanishing rest mass, coupled to Einstein's  equations
\begin{equation}
\label{I-Einst-equ}
\hat{R}_{\mu\nu} - \frac{1}{2}\,\hat{R}\,\hat{g}_{\mu\nu} + \lambda\,\hat{g}_{\mu\nu}
= \kappa\,\hat{T}_{\mu\nu},
\end{equation}
with cosmological constant $\lambda > 0$. The precise properties of the equations will depend there very much on the specific nature of the chosen matter field and each of them will need a special analysis. We shall concentrate here on the case of a non-linear scalar field $\phi$ which obeys an equation of the form 
\begin{equation}
\label{I-K-G-equ}
\hat{\nabla}_{\mu}\hat{\nabla}^{\mu}\phi - (m^2\,\phi + V'(\phi)) = 0,
\end{equation}
with energy momentum tensor
\begin{equation}
\label{I-K-G-em-tensor}
\hat{T}_{\mu\nu} = \hat{\nabla}_{\mu}\phi\,\hat{\nabla}_{\nu}\phi
- \left(\frac{1}{2}\,(\hat{\nabla}_{\rho}\phi\,\hat{\nabla}^{\rho}\phi + m^2\,\phi^2)
+ V(\phi)\right)\hat{g}_{\mu\nu},
\end{equation}
and a potential $V(\phi) = \mu\,\phi^3 + \phi^4\,U(\phi)$ with an arbitrary real coefficient $\mu$ and a smooth real-valued function $U$.

The future stability of such systems has been studied under quite general assumptions by Hans 
Ringstr\"om  \cite{ringstroem:2008}. His analysis will not be followed here. 
Employing a general wave gauge, he made a skillful choice of gauge source functions which allowed him to obtain estimates that conveniently control the long time behaviour of the fields, which is at the focus of his work. A sharp statement about the asymptotic behaviour of the fields, however, which is our main interest here, requires an optimal control (whatever that may mean in the end) on the fields as well as on the coordinates with respect to  which the behaviour of the fields is expressed. It is not easy to see whether Rinstr\"om's coordinates admit a precise description of the asymptotics of the gravitational and the matter fields.

To see precisely what may go wrong at the prospective conformal boundary we study instead in section \ref{cfe} the conformal field equations for general matter fields along the lines indicated in \cite{friedrich:1991}  and then specialize in section \ref{massive-fields} the matter model to that of a non-linear massive scalar field. As expected, it turns out that the conformal equations for the scalar field as well as those for the geometric background fields are in general strongly singular; if the equations are written so that the principal parts are well behaved, independent of the sign of the conformal factor $\Omega$, there occur in general in the lower order terms factors of the form $\Omega^{-k}$, $k = 1, 2$. These will blow up precisely at the set where the conformal factor $\Omega$ approaches zero, i.e. at  the prospective location of 
the conformal boundary.
One type of singularity is associated with the coefficient $\mu$. Assuming that $\mu = 0$, we get rid of it. The remaining  singularities are related to the mass $m$.
It turns out  that the singular terms occur in the scalar field equation as well as in the geometric background equations always
in the form
$\left(m^2 - \frac{2}{3}\,\lambda\right)\Omega^{-k}$, $ k = 1, 2$. Somewhat unexpected, the complete system of  conformal field equations will thus be  regular if the single condition 
\[
(*) \quad \quad \quad m^2 = \frac{2}{3}\,\lambda,\quad \quad \quad 
\]
is imposed. We note that it  can be non-trivially satisfied for real $m \neq 0$ only with the present sign of the cosmological constant. In the rest of the introduction  we shall assume this condition to be satisfied.

The construction of initial data for the system (\ref{I-Einst-equ}), (\ref{I-K-G-equ}), 
(\ref{I-K-G-em-tensor}) has been studied under fairly general assumptions  in 
\cite{choque-bruhat-isenberg-pollack:2007}. From the discussion above it is clear, however,  that not all of these data will develop into solutions that admit a smooth conformal boundary. To get an insight into the class of solutions which do admit such boundaries, we analyze  in section \ref{constraints} the constraints induced by the conformal field equations on the set $\{\Omega = 0\}$, assuming it to be given as a smooth compact 
$3$-manifold in the conformally extended solution space-time. 
As observed already in the vacuum case (\cite{friedrich:1986a}), the Hamiltonian constraint drops out. 
This leads to a considerable simplification, the data can be prescribed freely up to solving a linear system.

In a suitable gauge the conformal field equations for the coupled system induce again hyperbolic evolution equations that preserve the constraints and the gauge conditions. 
They can be used to determine forward and backward time developments of the data on 
$\{\Omega = \,0\}$ which provide  away from $\{\Omega = \,0\}$ solutions to the system 
(\ref{I-Einst-equ}), (\ref{I-K-G-equ}), (\ref{I-K-G-em-tensor}).
Denote by ${\cal A}_S$ the Cauchy data induced on the Cauchy hypersurfaces of these solutions
and  by  ${\cal D}_S$ all Cauchy data for the system (\ref{I-Einst-equ}), (\ref{I-K-G-equ}), (\ref{I-K-G-em-tensor}). It follows then as before, that the set ${\cal A}_S$ is open in ${\cal D}_S$. In fact, it follows with the observations mentioned above that
the solutions, their asymptotic behaviour, as well as their smooth extensibility across the conformal boundary are not only stable under perturbations of the scalar field and the geometric background fields but the perturbations may also involve zero rest-mass fields.

\vspace{.1cm}

That there has to be observed  
a  specific relation between the mass $m$ and the cosmological constant $\lambda$ could give rise to worries if each of these quantities already had a specific meaning of its own independent of the other one.  So far, however, the matter field and its mass have no specific interpretation and if they are  given one, the relation to the cosmological may even acquire some predictive power (`a relation between the cosmological constant and the dark matter' ?).  All this depends on the choice of the matter field and  the role assigned to it in a space-time model. In this context it should be emphasized that  the discussion in this article has been restricted to the scalar field only for the purpose of illustration,  other fields could  be considered as well.

To get an idea of the order of magnitude of the mass considered here we use the value  
$\lambda \approx 1,7 \times 10^{-121}$ in Planck units given in \cite{barrow-shaw:2011} 
(ignoring the fact that the cosmological model underlying the derivation of this value is different from the one referred to in the beginning of this article). Replacing $m$ in the equation above  by $\frac{m\,c}{\hbar}$ 
and converting units we find the exceedingly small mass $m \approx 4 \times 10^{-33}\,eV/c^2$.
It is interesting  to note that  in a study concerned  with the recent acceleration of the universe  
 Leonard Parker and Alpan Raval were led to consider,  by a completely different reasoning,  masses of a similar order of magnitude \cite{parker-raval-1999}.

\vspace{.1cm}

This article  focusses on the unexpected fact that the conformal equations can be regular at the
conformal boundaries. This does not mean that cases of the Einstein-scalar field system for which  condition $(*)$ is violated cannot be of interest.  On the contrary, it would be most interesting to understand the significance  of condition $(*)$ with respect to the asymptotic behavior of solutions in this set and to see
whether the system (\ref{I-Einst-equ}), (\ref{I-K-G-equ}), (\ref{I-K-G-em-tensor})
admits a range of masses for which the notion of  
crossover surface can be generalized and the conformal structure can be extended in a unique way. 
After all, it can be expected that transitions from exponentially expanding to big bang phases  are accompanied with losses of smoothness and that the insistence on too strict smoothness requirements may obstruct a modeling of such transitions.

\vspace{.1cm}

In section \ref{hom-sols} we discuss whether the evolution of massive field across the crossover surfaces may allow us to get some insight into  the problem of this  `phase transition'. 
To simplify matters, we set $\kappa = 1$ (or absorb it into the scalar field) and assume a scaling by a constant overall factor so that $\lambda = 3$ whence $m = \sqrt{2}$.
Then we consider spatially homogeneous solutions with a linear massive scalar field so that the physical fields are of the form
\[
\hat{g} = - dt^2 + f^2\,k, \quad  \phi = \phi(t) \quad \mbox{on} \quad 
\mathbb{R} \times S,
\]
where $f = f(t) > 0$ and $k$ denotes a Riemannian metric with constant curvature  
$R_{abcd}[k] = 2\,\epsilon\,k_{a[c}\,k_{d]b}$, $\epsilon = 1, 0, -1$.
In a convenient conformal and coordinate gauge the conformal metric then takes the form 
\[
g = - d\tau^2 + k,
\]
and the conformal field equations reduce to a regular system of ODE's of second order 
for $\Omega$ and the rescaled matter field $\psi = \Omega^{-1}\,\phi$ 
and a constraint which involves 
$\Omega$, $\psi$ and their derivatives of first order. We consider solutions determined by the backward development of data on $\{\tau = 0\} = \{\Omega = 0\}$. The constraint is satisfied there if 
$\Omega'(0) = -1$ while the data $\psi(0)$, $\psi'(0)$ can be prescribed freely. 

The most  interesting case  $\epsilon = 1$  is discussed in some detail.
If $\psi(0)$, $\psi'(0)$ are chosen to vanish, the solution is  given by $\Omega_{dS} = - \sin \tau$, $\psi_{dS} = 0$. 
Its  restriction to the intervall $- \pi < \tau < 0$, is conformal to the de Sitter space-time. 
The stability properties of  $\Omega_{dS} $, $\psi_{dS}$ then ensure that  there exists a large set of smooth solutions $\Omega$, $\psi \not \equiv 0$ so that
$\Omega(\tau_z) = \Omega(0) = 0$ for some $\tau_z < 0$ and $\Omega > 0$ in the interval
$]\tau_z, 0[$, in which it assumes its absolute maximum value $\Omega_m$ at a point 
$\tau_m$. 
The corresponding physical solutions can be thought of as arising from asymptotic  data on the `crossover surface' $\{\tau = \tau_z\}$, developing a `waist' of volume $\Omega_m^{-3}\,Vol(\mathbb{S}^3)$  at $\tau_m$,  and then expanding exponentially until they  approach the next crossover surface  at $\{\tau = 0\}$. We denote the set of these solutions by ${\cal B}$. All solutions  in ${\cal B}$ are non-linearly stable  under generic perturbations involving  the scalar field,  the geometric fields, and zero-rest mass fields.

There is a solution not belonging to ${\cal B}$ which is of particular interest in our context. It is given by $\Omega_* = - \tau$, $\psi_* = \sqrt{2}$. Its restriction to the domain where $\Omega > 0$ defines a physical field that  is given in terms of the coordinate $t = - \log (- \tau)$ by
\[
\tilde{g} = - dt^2 + e^{2t}\,k,
\quad \phi = \sqrt{2}\,e^{-t}.
\]
As $t \rightarrow - \infty$ the matter field $\phi$ grows unboundedly while it decays and  the metric  shows  a de Sitter-type expansion behaviour as $t \rightarrow \infty$. 
If the solution $\Omega_*$, $\psi_*$ could be approximated on any given interval of the form
$[z, 0]$, $z < 0$, by solutions
in ${\cal B}$ there will exist solutions with an arbitrarily narrow waist. The restriction of such solutions to the range $]\tau_m, 0[$ would, from the point of view of observational data, hardly be distinguishable from solutions which start with a big bang and then expand exponentially. No attempt is made in this article to decide about
this question because it  requires a detailed analysis of the solution space.

\section{The conformal field equations.}
\label{cfe}

We consider a 4-dimensional manifold $M$ with smooth boundary ${\cal J}$ and interior
$\hat{M} = M \setminus {\cal J}$. Let $\hat{g}$ and $g$ denote Lorentz metrics on $\hat{M}$ and $M$ respectively which satisfy
\[
g_{\mu\nu} = \Omega^2\,\hat{g}_{\mu\nu}  \quad \mbox{on} \quad \hat{M},
\]
with a conformal factor  $\Omega$ that is given by a smooth function on $M$ such that
\[
 \Omega > 0  \quad \mbox{on} \quad \hat{M}, \quad \Omega = 0, \,\,d\Omega \neq 0
\quad \mbox{on} \quad {\cal J}.
\]
In the following ${\cal J}$ will be thought of as being  space-like with respect to $g$, though in the end this will be a consequence of the field equations (cf. (\ref{dOmega-on-scri})).
It is assumed that $\hat{g}$ satisfies Einstein's field equations
\begin{equation}
\label{Einst-equ}
\hat{R}_{\mu\nu} - \frac{1}{2}\,\hat{R}\,\hat{g}_{\mu\nu} + \lambda\,\hat{g}_{\mu\nu}
= \kappa\,\hat{T}_{\mu\nu}
\end{equation}
with cosmological constant $\lambda > 0$. The matter fields will be specified later.

To formulate these equations in terms of the conformal metric $g$, the conformally transformed matter fields, and a number of fields derived from them,
we note the contraction
\begin{equation}
\label{contr-Einst-equ}
- \hat{R} + 4\,\lambda = \kappa\,\hat{T}.
\end{equation}
and use the general conformal transformation relation
\begin{equation}
\label{Ric-ten-transf}
R_{\mu\nu} + \frac{2}{\Omega}\,\nabla_{\mu}\,\nabla_{\nu}\Omega
+ \left(
\frac{1}{\Omega}\,\nabla_{\rho}\,\nabla^{\rho}\Omega
- \frac{3}{\Omega^2}\,\nabla_{\rho}\Omega\,\nabla^{\rho}\Omega
\right)g_{\mu\nu} = \hat{R}_{\mu\nu},
\end{equation}
and its trace
\begin{equation}
\label{Ric-scal-transf}
R + \frac{6}{\Omega}\,\nabla_{\rho}\,\nabla^{\rho}\Omega
- \frac{12}{\Omega^2}\,\nabla_{\rho}\Omega\,\nabla^{\rho}\Omega =
\frac{1}{\Omega^2}\,\hat{R},
\end{equation}
where the covariant derivative operator $\nabla$ and the index operations on the left hand sides refer to the metric $g$. 
With the definition
\begin{equation}
\label{s-def}
s \equiv \frac{1}{4}\,\nabla_{\rho}\,\nabla^{\rho}\Omega + \frac{1}{24}\,\Omega\,R,
\end{equation}
equation ({\ref{Ric-scal-transf}) takes the form
\begin{equation}
\label{B-Ric-scal-transf}
6\,\Omega\,s - 3\,\nabla_{\rho}\Omega\,\nabla^{\rho}\Omega = \frac{1}{4}\,\hat{R},
\end{equation}
and equation (\ref{Ric-ten-transf}) can be rewritten 
\begin{equation}
\label{B-Ric-ten-transf}
\nabla_{\mu}\,\nabla_{\nu}\Omega = - \,\Omega\,L_{\mu\nu} + s\,g_{\mu\nu}
+ \Omega\,\hat{S}_{\mu\nu},
\end{equation}
where 
\[
L_{\mu\nu} = \frac{1}{2}\,\left(R_{\mu\nu} - \frac{1}{6}\,R\,g_{\mu\nu}\right)
\quad \mbox{and} \quad 
\hat{S}_{\mu\nu}  = \frac{1}{2}\,\left(\hat{R}_{\mu\nu}  
- \frac{1}{4}\,\hat{R}\,\hat{g}_{\mu\nu}\right),
\]
denote the Schouten tensor of $g$ and (half of) the trace-free part of the Ricci tensor 
of $\hat{g}$ respectively.
To derive a differential equation for $s$ we observe that the general conformal transformation relations for covariant derivatives and derived tensor fields and the Bianchi identity  imply
\[
g^{\rho \nu}\,\nabla_{\rho}\,\hat{S}_{\nu \mu} =
\frac{1}{\Omega^2}\,\hat{g}^{\rho \nu}\,\hat{\nabla}_{\rho}\,\hat{S}_{\nu \mu} 
+ \frac{2}{\Omega}\,\nabla^{\rho}\,\Omega\,  \hat{S}_{\rho \mu}
=
\frac{1}{8\,\Omega^2}\,\hat{\nabla}_{\mu}\,\hat{R}
+ \frac{2}{\Omega}\,\nabla^{\rho}\,\Omega\,  \hat{S}_{\rho \mu}. 
\]
Applying a derivative to (\ref{B-Ric-ten-transf}), commuting on the left hand side covariant derivatives, and performing a contraction then gives
\begin{equation}
\label{s-equ}
\nabla_{\mu}\,s = - \,\nabla^{\rho}\Omega\,L_{\rho\mu} 
+ \nabla^{\rho}\Omega\,\hat{S}_{\rho\nu}
+ \frac{1}{24\,\Omega}\,\hat{\nabla}_{\mu}\,\hat{R}.
\end{equation}
With the decomposition 
\[
R_{\mu \rho \nu \lambda} = C_{\mu \rho \nu \lambda}
+ 2\left\{g_{\mu [\nu}\,L_{\lambda]\rho}
+L_{\mu [\nu }\,g_{\lambda]\rho}\right\},
\]
of the curvature tensor into the conformal Weyl tensor and the Schouten tensor 
the once contracted Bianchi identity for the curvature tensor of $g$ can be written
\begin{equation}
\label{contr-Bianchi}
\nabla_{\mu}C^{\mu}\,_{\rho \nu \lambda} = 2\,\nabla_{[\nu}\,L_{\lambda] \rho},
\end{equation} 
and  the analogue for $\hat{g}$ reads
\begin{equation}
\label{hat-contr-Bianchi}
\hat{\nabla}_{\mu}\hat{C}^{\mu}\,_{\rho \nu \lambda} 
= 2\,\hat{\nabla}_{[\nu}\,\hat{L}_{\lambda] \rho}.
\end{equation}
With the conformal covariance relations
\[
C^{\mu}\,_{\rho \nu \lambda} = \hat{C}^{\mu}\,_{\rho \nu \lambda},
\quad \quad
\nabla_{\mu}(\Omega^{-1}\,C^{\mu}\,_{\rho \nu \lambda} ) = 
\Omega^{-1}\,\hat{\nabla}_{\mu}\,\hat{C}^{\mu}\,_{\rho \nu \lambda}, 
\] 
and the definition
\[
W^{\mu}\,_{\rho \nu \lambda}  \equiv \Omega^{-1}\,C^{\mu}\,_{\rho \nu \lambda},
\]
equation (\ref{hat-contr-Bianchi}) can be written
\begin{equation}
\label{W-equ}
\nabla_{\mu}\,W^{\mu}\,_{\rho \nu \lambda} 
= 2\,\Omega^{-1}\,\hat{[\nabla}_{\nu}\,\hat{L}_{\lambda] \rho},
\end{equation}
while (\ref{contr-Bianchi}) reads
\begin{equation}
\label{L-equ}
\nabla_{\nu}\,L_{\lambda \rho} 
- \nabla_{\lambda}\,L_{\nu \rho} = 
\nabla_{\mu}\Omega\,\,W^{\mu}\,_{\rho \nu \lambda} 
+ 2\,\hat{\nabla}_{[\nu}\,\hat{L}_{\lambda] \rho}.
\end{equation}
Taking  now into account the the field equations (\ref{Einst-equ}), we get the equations above in the form 

\begin{equation}
\label{mat-B-Ric-scal-transf}
6\,\Omega\,s - 3\,\nabla_{\rho}\Omega\,\nabla^{\rho}\Omega - 
\lambda = - \frac{\kappa}{4}\,\hat{T},
\end{equation}

\begin{equation}
\label{mat-B-Ric-ten-transf}
\nabla_{\mu}\,\nabla_{\nu}\Omega + \,\Omega\,L_{\mu\nu} - s\,g_{\mu\nu}
= \frac{\kappa}{2}\,\Omega\,T^*_{\mu\nu},
\end{equation}

\begin{equation}
\label{mat-s-equ}
\nabla_{\mu}\,s + \nabla^{\rho}\Omega\,L_{\rho\mu} 
= \frac{\kappa}{2}\,\nabla^{\rho}\Omega\,T^*_{\rho \mu}
- \frac{\kappa}{24\,\Omega}\,\hat{\nabla}_{\mu}\,\hat{T},
\end{equation}

\begin{equation}
\label{mat-L-equ}
\nabla_{\nu}\,L_{\lambda \rho} 
- \nabla_{\lambda}\,L_{\nu \rho} - 
\nabla_{\mu}\Omega\,\,W^{\mu}\,_{\rho \nu \lambda} 
= 2\,\hat{\nabla}_{[\nu}\,\hat{L}_{\lambda] \rho},
\end{equation}

\begin{equation}
\label{mat-W-equ}
\nabla_{\mu}\,W^{\mu}\,_{\rho \nu \lambda} 
= 2\,\Omega^{-1}\,\hat{\nabla}_{[\nu}\,\hat{L}_{\lambda] \rho},
\end{equation}
with
\[
T^*_{\rho \mu} 
= \hat{T}_{\rho \mu} - \frac{1}{4}\,\hat{T}\,\hat{g}_{\rho \mu}, \quad \quad
\hat{\nabla}_{\rho} \hat{L}_{\mu \nu} = \frac{\kappa}{2}\,\hat{\nabla}_{\rho} \left(\hat{T}_{\mu \nu} 
- \frac{1}{3}\,\hat{T}\,\hat{g}_{\mu \nu} 
\right).
\]

The first of these equations may be considered as a constraint which will be satisfied if the initial data are arranged accordingly:
{\it If (\ref{mat-B-Ric-scal-transf})  holds at a point $p$ and (\ref{mat-B-Ric-ten-transf}),
(\ref{mat-s-equ}) are satisfied on a connected neighbourhood $U$ of $p$, then 
 (\ref{mat-B-Ric-scal-transf}) holds on $U$.}
In fact, a direct calculation using (\ref{mat-B-Ric-ten-transf}),
(\ref{mat-s-equ}) implies that 
\[
\nabla_{\mu}(6\,\Omega\,s - 3\,\nabla_{\rho}\Omega\,\nabla^{\rho}\Omega  
- \lambda + \frac{\kappa}{4}\,\hat{T}) = 0.
\]
The equations above for the tensorial inknowns
$\Omega$, $s$, $L_{\mu \nu}$, $W^{\mu}\,_{\rho \nu \lambda}$,
have to be combined with equations which determine the metric and the connection. One possibility to do this is to write  the structural equations as equations for the unknowns
$e^{\mu}\,_k$, $\Gamma_i\,^k\,_j$,
where the first set of fields are the coefficients of a $g$-orthonormal frame field
$e_k = e^{\mu}\,_k\,\partial_{x^{\mu}}$ with respect to a coordinate system $x^{\mu}$ 
so that $g(e_i, e_j) = g_{\mu\nu}\,e^{\mu}\,_i\,e^{\nu}\,_j  = \eta_{ij}$
and the second set of fields are the associated connection coefficients $\Gamma_i\,^k\,_j$ defined by
$\nabla_i\,e_j = \Gamma_i\,^k\,_j\,e_k$ with $\nabla_i = \nabla_{e_i}$, which satisfy
$\Gamma_{ijk} = - \Gamma_{ikj}$ where $\Gamma_{ijk} = \Gamma_i\,^l\,_k\,\eta_{lj}$. 
In terms of these unknowns the structural equations  take the form
of the
{\it torsion-free condition} 
\begin{equation}
\label{torsion-free condition}
e^{\mu}\,_{i,\,\nu}\,e^{\nu}\,_{j}
 - e^{\mu}\,_{j,\,\nu}\,e^{\nu}\,_{i} 
= (\Gamma_{j}\,^{k}\,_{i} - \Gamma_{i}\,^{k}\,_{j})\,e^{\mu}\,_{k},
\end{equation}
and the {\it Ricci identity}
\begin{equation}
\label{Ricci identity}
\Gamma_l\,^i\,_{j,\,\mu}\,e^{\mu}\,_k - \Gamma_k\,^i\,_{j,\,\mu}\,e^{\mu}\,_l
+ 2\,\Gamma_{[k}\,^{i\,p}\,\Gamma_{l]pj}
- 2\,\Gamma_{[k}\,^p\,_{l]}\,\Gamma_p\,^i\,_j
\end{equation}
\[
= \Omega\,W^i\,_{jkl}
+ 2\,\{g^i\,_{[k}\,L_{l] j} + L^i\,_{ [k}\,g_{l] j}\}.
\]
If equations 
(\ref{mat-B-Ric-scal-transf}) to (\ref{mat-W-equ}) are expressed in terms of the frame and combined with the strutural equations, they are equivalent to Einstein's vacuum equations where
$\Omega \neq 0$. \\

The vacuum case is characterized by vanishing right hand sides of equations (\ref{mat-B-Ric-scal-transf}) to (\ref{mat-W-equ}). If the resulting system is written with respect to a suitable choice of coordinates and frame field, and if the conformal factor is controlled by specifying the Ricci scalar as a function of the coordinates (which can locally be prescribed in an arbitrary way), the combined system implies equations {\it which are hyperbolic even where $\Omega$ changes sign}. Moreover, they preserve the constraints and the gauge conditions.
Dicussions of this fact, giving various versions of hyperbolic systems, can be found in 
\cite{friedrich:asymp-sym-hyp:1981b}, \cite{friedrich:1986a}, \cite{friedrich:1986b}.
The case of zero rest-mass fields for which the energy momentum tensor is trace free is similar and has been discussed in \cite{friedrich:1991}. The details will not be reproduced  here.

In the following we will be interested in fields with non-vanishing rest-mass. The further analysis  depends very much on the specific behaviour of the 
matter fields and the associated energy momentum tensor under conformal rescalings.

\section{The non-linear massive scalar field.}
\label{massive-fields}

In the following we consider a scalar field $\phi$ that satisfies an equation of the form
\begin{equation}
\label{K-G-equ}
\hat{\nabla}_{\mu}\hat{\nabla}^{\mu}\phi - (m^2\,\phi + V'(\phi)) = 0,
\end{equation}
with energy momentum tensor
\begin{equation}
\label{K-G-em-tensor}
\hat{T}_{\mu\nu} = \hat{\nabla}_{\mu}\phi\,\hat{\nabla}_{\nu}\phi
- \left(\frac{1}{2}\,(\hat{\nabla}_{\rho}\phi\,\hat{\nabla}^{\rho}\phi + m^2\,\phi^2)
+ V(\phi)\right)\hat{g}_{\mu\nu},
\end{equation}
and a potential of the form 
\begin{equation}
\label{potential}
V(\phi) = \mu\,\phi^3 + \phi^4\,U(\phi),
\end{equation}
where $\mu$ is a real coefficient and $U$ a smooth real-valued function.
This form is assumed because we wish to discuss the cosmological constant, the mass term, and the constant coefficient   $\mu$  seperately and because $V'(0) = 0$ ensures that the coupled Einstein scalar field equations admit solutions with $\phi = 0$.

 In four dimensions holds for arbitrary smooth functions $\phi$ the transformation law  
\begin{equation}
\label{scal-trafo}
\left
(\Box_g - \frac{1}{6}\,R\right)[\Omega^{-1}\,\phi] 
= \Omega^{-3}\,\left(\Box_{\hat{g}} - \frac{1}{6}\,\hat{R}\right)[\phi]. 
\end{equation}
In terms of the new unknown
\[
\psi = \Omega^{-1}\,\phi,
\]
equation (\ref{K-G-equ}) takes then with (\ref{contr-Einst-equ})  
the form
\begin{equation}
\label{conf-K-G-equ}
\left
(\Box_g - \frac{1}{6}\,R\right)[\psi] 
= \Omega^{-2}\left(m^2 -  \frac{2}{3}\,\lambda + \frac{\kappa}{6}\,\hat{T}\right)\psi  + \Omega^{-3}\,V'(\Omega\,\psi),
\end{equation}
where 
\[
\Omega^{-3}\,V'(\Omega\,\psi) = 3\,\mu\,\Omega^{-1}\,\psi^2 
+ 4\,\psi^3\,U(\Omega\,\psi) + \Omega\,\psi^4\,U'(\Omega\,\psi).
\]
To ensure the regularity of this term where $\Omega \rightarrow 0$ it will be assumed in the following that
\[
\mu = 0.
\]
The trace of the energy momentum tensor (\ref{K-G-em-tensor}) now reads
\begin{equation}
\label{trace-e-m-tensor}
\hat{T} = 
- \Omega^2\,\left\{\nabla_{\rho}(\Omega\,\psi)\,\nabla^{\rho}(\Omega\,\psi) + 2\,m^2\,\psi^2
+ 4\, \Omega^{-2}\,V(\Omega\,\psi)\right\},
\end{equation}
so that 
\begin{equation}
\label{rescaled-trace-e-m-tensor}
T_* \equiv \Omega^{-2}\hat{T} = 
- \nabla_{\rho}(\Omega\,\psi)\,\nabla^{\rho}(\Omega\,\psi) - 2\,m^2\,\psi^2
- 4\, \Omega^{-2}\,V(\Omega\,\psi),
\end{equation}
is well behaved as $\Omega \rightarrow 0$. The relations  (\ref{contr-Einst-equ}) and (\ref{B-Ric-scal-transf}) imply
\begin{equation}
\label{grad-Om-squared}
\nabla_{\rho}\Omega\,\nabla^{\rho}\Omega =  - \frac{\lambda}{3}
+ 2\,\Omega\,s + \frac{\kappa}{12}\,\Omega^2\,T_*,   
\end{equation}
and thus 
\begin{equation}
\label{T_*-behaviour}
T_* 
= \left(\frac{\lambda}{3} - 2\,m^2\right)\psi^2 
- 2\,\Omega\,(s\,\psi^2 + \psi\,\nabla_{\pi}\Omega\,\nabla^{\pi}\psi)
\quad \quad \quad \quad \quad
\end{equation}
\[
\quad \quad \quad \quad \quad
- \Omega^2\,\left(\frac{\kappa}{12}\,\psi^2\,T_* + \nabla_{\pi}\psi\,\nabla^{\pi}\psi\right)
- 4\,\Omega^{-2}\,V(\Omega\,\psi).
\]
From this we get  
\begin{equation}
\label{final-T_*-behaviour}
T_* = \left(1 + \frac{\kappa}{12}\,\psi^2\, \Omega^2\right)^{-1}
\left\{\left(\frac{\lambda}{3} - 2\,m^2\right)\psi^2 
- 2\,\Omega\,(s\,\psi^2 + \psi\,\nabla_{\pi}\Omega\,\nabla^{\pi}\psi) 
\right.
\end{equation}
\[
\left.
- \frac{}{}\Omega^2\,\nabla_{\pi}\psi\,\nabla^{\pi}\psi
- 4\,\Omega^{-2}\,V(\Omega\,\psi)\right\},
\]
which is  a smooth function for all (real) values of the unkowns $\psi$, $\nabla_{\mu}\psi$, $\Omega$, and $s$. In the following calculations  it will be convenient, however,  to use the equation in the form (\ref{T_*-behaviour}).
It follows 
\begin{equation}
\label{B-conf-K-G-equ}
\Box_g \psi - \frac{1}{6}\,R\,\psi 
= \Omega^{-2}\left(m^2 -  \frac{2}{3}\,\lambda\right)\psi  
+ \frac{\kappa}{6}\,T_* \psi  + 4\,\psi^3\,U(\Omega\,\psi) + \Omega\,\psi^4\,U'(\Omega\,\psi).
\end{equation}
We note that  if the background fields are given this is a semi-linear equation for $\psi$ whose right hand side depends via  $T_*$ also on the derivative $\nabla_{\mu}\psi$.
Its most conspicuous feature, however, is the first term on the right hand side, which is singular where
$\Omega \rightarrow 0$. 

\vspace{.2cm}
\noindent
We are in a position now to state our main result.

\begin{theorem}
\label{main-result}
Consider the energy momentum tensor given by (\ref{K-G-em-tensor}), a potential (\ref{potential})
with $\mu = 0$, and the coupled system of equations 
(\ref{mat-B-Ric-scal-transf}),
(\ref{mat-B-Ric-ten-transf}), (\ref{mat-s-equ}), (\ref{mat-L-equ}), (\ref{mat-W-equ}),
(\ref{torsion-free condition}), (\ref{Ricci identity}), 
(\ref{B-conf-K-G-equ})
for the unknowns
\begin{equation}
\label{unknowns}
e^{\mu}\,_k, \quad \Gamma_i\,^k\,_j, \quad \Omega, \quad s, \quad L_{\mu \nu}, \quad W^{\mu}\,_{\rho \nu \lambda}, \quad \psi.
\end{equation} 
If and only if the single condition 
\begin{equation}
\label{scal-field-reg-cond}
m^2 =  \frac{2}{3}\,\lambda,
\end{equation}
is satisfied this system
is regular in the sense that on the right hand sides of the equations no terms of the form 
$\Omega^{-k}$, $k > 0$, occur and the right hand sides are in fact smooth function of the unknowns.

\end{theorem}

\noindent
{\bf Remarks}:

We note that the condition above can be satisfied with real $m$ only with the de Sitter-type  sign of the cosmological constant.

Because some of the equations involve derivatives of the energy momentum tensor they contain derivatives of $\psi$ of second order. Applying a derivative to (\ref{B-conf-K-G-equ}) and commuting operators one obtains a wave equation for $\nabla_{\mu}\psi$ and thus altogether a quasi-linear, overdetermined system of equations for the unknowns (\ref{unknowns}) and $\nabla_{\mu}\psi$. 
After fixing a suitable gauge one can extract from the complete set of equations a hyperbolic evolution system which preserves the gauge conditions and the constraints. Since various versions of this procedure have been discussed at lenght in the references given above we shall not go into the details here.

It should be noted that the compactness of the manifold $S$ plays no role in this result.

\vspace{.3cm}

\noindent
{\bf Proof of  Theorem \ref{main-result}}:
It follows immediately that (\ref{scal-field-reg-cond}) renders equation (\ref{B-conf-K-G-equ})
regular. Equations (\ref{torsion-free condition}), (\ref{Ricci identity}) are obviously regular.
We discuss the nature of the singularity of the remaining equations. The trace free part of the energy momentum tensor  takes the form
\begin{equation}
\label{trace-free-part-e-m-tensor}
T^*_{\mu\nu} 
= \nabla_{\mu}(\Omega\psi)\,\nabla_{\nu}(\Omega\,\psi)
- \frac{1}{4}\,\nabla_{\rho}(\Omega\psi)\,\nabla^{\rho}(\Omega\psi)\,\,g_{\mu\nu},
\end{equation} 
\[
= \psi^2(\nabla_{\mu}\Omega\,\nabla_{\nu}\Omega 
- \frac{1}{4}\,\nabla_{\pi}\Omega\,\nabla^{\pi}\Omega\,g_{\mu\nu})
\]
\[
+ 2\,\Omega\,\psi\,(\nabla_{(\mu}\Omega\,\nabla_{\nu)}\psi 
- \frac{1}{4}\,\nabla_{\pi}\Omega\,\nabla^{\pi}\psi\,g_{\mu\nu})
\]
\[
+ \Omega^2\,(\nabla_{\mu}\psi\,\nabla_{\nu}\psi 
- \frac{1}{4}\,\nabla_{\pi}\psi\,\nabla^{\pi}\psi\,g_{\mu\nu}),
\]
and is thus regular. It follows that equation (\ref{mat-B-Ric-scal-transf}), 
given now by (\ref{grad-Om-squared}) and equations (\ref{mat-B-Ric-ten-transf}), 
and (\ref{mat-s-equ})  with 
$\hat{\nabla}_{\mu}\,\hat{T}  = \nabla_{\mu}\,(\Omega^2\,T_*)$, 
are  regular as $\Omega  \rightarrow 0$. 
Critical are equations 
(\ref{mat-L-equ}), (\ref{mat-W-equ}).
With the notation above we have 
\[
\hat{L}_{\mu\nu} = \frac{\kappa}{2}\,T^*_{\mu\nu} 
- \frac{\kappa}{24}\,\,\hat{T}\,\hat{g}_{\mu\nu}
+ \frac{\lambda}{6}\,\hat{g}_{\mu\nu}.
\]
While the first two terms on the right hand side are regular since
$\hat{T}\,\hat{g}_{\mu\nu} = T_*\,g_{\mu\nu}$, 
the last term is singular if it is expressed in terms of $g_{\mu\nu}$. However, this term is annihilated by the operator 
$\hat{\nabla}_{\rho}$ and it follows 
\[
\hat{\nabla}_{\rho} \hat{L}_{\mu \nu}
= \frac{\kappa}{2}\,\hat{\nabla}_{\rho} 
\,T^*_{\mu \nu} 
- \frac{\kappa}{24}\,\Omega^{-2}\,\nabla_{\rho}(\Omega^2\,T_*)\,g_{\mu \nu} 
\]
\[
= \frac{\kappa}{2}\,\nabla_{\rho} \,T^*_{\mu \nu} 
- \frac{\kappa}{24}\,\nabla_{\rho}T_*\,g_{\mu \nu} 
- \frac{\kappa}{12}\,\Omega^{-1}\,T_*\,\nabla_{\rho}\Omega\,g_{\mu \nu} 
\]
\[
+  \frac{\kappa}{2}\,\Omega^{-1}\,\left(3\,\nabla_{(\rho}\,\Omega\,T^*_{\mu) \nu} 
+ \nabla_{[\rho}\Omega\,T^*_{\mu] \nu} 
+ \nabla_{\nu}\Omega\,T^*_{\rho \mu} - g_{\rho \mu}\,T^*_{\nu \delta} \nabla^{\delta}\Omega
- g_{\nu \rho}\,T^*_{\mu \delta} \nabla^{\delta}\Omega\right),
\]
whence
\begin{equation}
\label{d-hat-L}
\hat{\nabla}_{[\rho} \hat{L}_{\mu] \nu}
= \frac{\kappa}{2}\left\{\nabla_{[\rho} \,T^*_{\mu] \nu} 
- \frac{1}{12}\,\nabla_{[\rho}T_*\,g_{\mu] \nu}
\right. 
\quad \quad \quad \quad \quad \quad \quad \quad 
\end{equation}
\[
\quad \quad \quad \quad \quad \quad \quad 
\left.
+ \Omega^{-1}\left(
\nabla_{[\rho}\Omega\,T^*_{\mu] \nu} 
+ \nabla^{\pi}\Omega\,T^*_{\pi [\rho} \,g_{\mu] \nu}\,
- \frac{1}{6}\,T_*\,\nabla_{[\rho}\Omega\,g_{\mu] \nu}\right)\right\}.
\]
Direct calculations using  (\ref{trace-free-part-e-m-tensor}),
 (\ref{mat-B-Ric-ten-transf}) and (\ref{grad-Om-squared}) give
\[
\nabla_{[\rho} \,T^*_{\mu] \nu} = 
\]
\[
\frac{\lambda}{6}\,\psi\,\nabla_{[\rho}\psi\,g_{\mu] \nu}
- \psi\,\nabla_{[\rho}\Omega\,\nabla_{\mu]}\psi\,\nabla_{\nu}\Omega
- \left(\frac{3}{2}\,s\,\psi^2 + \frac{1}{2}\,\psi\,\nabla_{\pi}\Omega\,\nabla^{\pi}\,\psi
\right)\,\nabla_{[\rho}\Omega\,g_{\mu] \nu}
\] 
\[
+ \Omega\left\{ 
- \frac{1}{2}\,\nabla_{\pi}\psi\,\nabla^{\pi}\psi\,\nabla_{[\rho}\Omega\,g_{\mu] \nu}
- \left(2\,s\,\psi + \frac{1}{2}\,\nabla_{\pi}\Omega\,\nabla^{\pi}\psi\right)
\nabla_{[\rho}\psi\,g_{\mu] \nu} \right.
\]
\[
+ \psi^2\left(\nabla_{[\rho}\Omega\,L_{\mu] \nu}
+ \frac{1}{2}\,\nabla^{\pi}\Omega\,L_{\pi[\rho}\,g_{\mu]\nu}\right)
- \psi\left(\nabla_{[\rho}\Omega\,\nabla_{\mu]} \nabla_{\nu}\psi
+ \frac{1}{2}\,\nabla^{\pi}\Omega\,\nabla_{\pi} \nabla_{[\rho}\psi\,g_{\mu] \nu}\right)
\]
\[
\left. + \nabla_{[\rho}\Omega\,\nabla_{\mu]}\psi\,\nabla_{\nu}\psi
- \frac{\kappa}{2}\psi^2\,\nabla_{[\rho}\Omega\,T^*_{\mu]\nu} 
- \frac{\kappa}{4}\psi^2\,\nabla^{\pi}\Omega\,T^*_{\pi [\rho}\,g_{\mu] \nu} 
\right\}
\]
\[
+ \Omega^2\left\{\psi\,\nabla_{[\rho}\psi\,L_{\mu]\nu} 
- \nabla_{[\rho}\psi\,\nabla_{\mu]}\nabla_{\nu}\psi 
-\frac{1}{2}\,\nabla^{\pi}\psi\,\nabla_{\pi}\nabla_{[\rho}\psi\,g_{\mu]\nu} 
\right.
\]
\[
\left.
-\frac{\kappa}{24}\,\psi\,T_*\,\nabla_{[\rho}\psi\,g_{\mu]\nu}
-\frac{\kappa}{2}\,\psi\,\nabla_{[\rho}\psi\,T^*_{\mu] \nu}
\right\},
\]

\[
- \frac{1}{12}\,\nabla_{[\rho}T_*\,g_{\mu] \nu} = 
\]
\[
\left(\frac{m^2}{3} - \frac{\lambda}{18}\right)\psi\,\nabla_{[\rho}\psi\,g_{\mu]\,\nu}
+ \frac{1}{6}\,(s\,\psi^2 
+ \psi\,\nabla_{\pi}\Omega\,\nabla^{\pi}\psi)\nabla_{[\rho}\Omega\,g_{\mu] \nu}
\]
\[
+ \Omega\left\{
\left(\frac{1}{2}\,s\,\psi + \frac{1}{6}\,\nabla_{\pi}\Omega\,\nabla^{\pi}\psi\right)\,\nabla_{[\rho}\psi\,g_{\mu]\nu}
+ \frac{1}{6}\,\nabla_{\pi}\psi\,\nabla^{\pi}\psi\,\nabla_{[\rho}\Omega\,g_{\mu]\nu}
\right.
\]
\[
\left. 
+ \frac{1}{6}\,\psi\,\nabla^{\pi}\Omega\,\nabla_{\pi}\nabla_{[\rho}\psi\,g_{\mu]\nu}
- \frac{1}{6}\,\psi^2\,\nabla^{\pi}\Omega\,L_{\pi[\rho}\,g_{\mu]\nu}
+ \frac{\kappa}{12}\,\psi^2\,\nabla^{\pi}\Omega\,T^*_{\pi[\rho}\,g_{\mu]\nu}
\right\}
\]
\[
+ \Omega^2\left\{
\frac{\kappa}{72}\,\psi\,T_*\,\nabla_{[\rho}\psi\,g_{\mu]\,\nu}
+  \frac{\kappa}{12}\,\psi\,\nabla^{\pi}\psi\,T^*_{\pi [\rho}\,g_{\mu] \nu}\right.
\]
\[
\left.
- \frac{1}{6}\,\psi\,\nabla^{\pi}\psi\,L_{\pi [\rho}\,g_{\mu] \nu}
+ \frac{1}{6}\,\nabla^{\pi}\psi\,\nabla_{\pi}\nabla_{[\rho}\psi\,g_{\mu] \nu}
 \right\},
\]
\[
- \frac{1}{3} \left\{\left(2\,\Omega^{-3}\,V(\Omega\,\psi) 
- \Omega^{-2}\,\psi\,V'(\Omega\,\psi)\right)\nabla_{[\rho}\Omega\,g_{\mu] \nu}
- \Omega^{-1}\,V'(\Omega\psi)\,\nabla_{[\rho}\psi\,g_{\mu] \nu}\right\}
\]

\[
\nabla_{[\rho}\,\Omega\,T^*_{\mu] \nu}  = \frac{\lambda}{12}\,\psi^2\,\nabla_{[\rho}\Omega\,g_{\mu] \nu}
\]
\[
+ \Omega \left\{- \frac{1}{2}\,(s\,\psi^2
+ \psi\,\nabla_{\pi}\Omega\,\nabla^{\pi}\,\psi)\,\nabla_{[\rho}\Omega\,g_{\mu] \nu}
+ \psi\,\nabla_{[\rho}\Omega\,\nabla_{\mu]}\psi\,\nabla_{\nu}\Omega
\right\} 
\]
\[
+ \Omega^2\,\left\{\nabla_{[\rho}\Omega\,\nabla_{\mu]}\psi\,\nabla_{\nu}\psi
- \left(\frac{1}{4}\,\nabla_{\pi}\psi\,\nabla^{\pi}\psi
+ \frac{\kappa}{48}\,\psi^2\,T_*\right)\nabla_{[\rho}\Omega\,g_{\mu] \nu}
\right\},
\]

\[
\nabla^{\pi}\Omega\,T^*_{\pi [\rho}\,g_{\mu] \nu} =
\]
\[
- \frac{\lambda}{4}\,\psi^2\,\nabla_{[\rho}\Omega\,g_{\mu] \nu}
+ \Omega \left\{
\left(\frac{3}{2}\,s\,\psi^2 + \frac{1}{2}\,\psi\,\nabla_{\pi}\Omega\,\nabla^{\pi}\psi
\right)\nabla_{[\rho}\Omega\,g_{\mu] \nu} 
- \frac{\lambda}{3}\,\psi\,\nabla_{[\rho}\psi\,g_{\mu] \nu}
\right\}
\]
\[
+ \Omega^2\left\{\left(2\,s\,\psi + \nabla_{\pi}\Omega\,\nabla^{\pi}\psi
+ \frac{\kappa}{12}\Omega\,\psi\,T_*\right)\nabla_{[\rho}\psi\,g_{\mu] \nu}
+ 
\left(\frac{\kappa}{16}\,\psi^2\,T_* - \frac{1}{4}\,\nabla_{\pi}\psi\,\nabla^{\pi}\psi
\right)\nabla_{[\rho}\Omega\,g_{\mu] \nu}
\right\},
\]

\[
- \frac{1}{6}\,T_*\,\nabla_{[\rho}\Omega\,g_{\mu] \nu} = 
\]
\[
\left( \frac{m^2}{3}  - \frac{\lambda}{18}\right)\psi^2\,\nabla_{[\rho}\Omega\,g_{\mu] \nu}
+ \Omega\left(\frac{1}{3}\,s\,\psi^2 
+ \frac{1}{3}\,\psi\,\nabla_{\pi}\Omega\,\nabla^{\pi}\psi
\right)\nabla_{[\rho}\Omega\,g_{\mu] \nu}
\]
\[
+ \Omega^2\left(\frac{1}{6}\,\nabla_{\pi}\psi\,\nabla^{\pi}\psi
+ \frac{\kappa}{72}\,\psi^2\,T_*
\right)\nabla_{[\rho}\Omega\,g_{\mu] \nu}
+ \frac{2}{3}\,\Omega^{-2}\,V(\Omega\,\psi)\,\nabla_{[\rho}\Omega\,g_{\mu] \nu},
\]
whence
\[
\frac{1}{\Omega}\left(\nabla_{[\rho}\Omega\,T^*_{\mu] \nu} 
+ \nabla^{\pi}\Omega\,T^*_{\pi [\rho}\,g_{\mu] \nu} 
- \frac{1}{6}\,T_*\,\nabla_{[\rho}\Omega\,g_{\mu] \nu}\right) 
 = 
 \]
 \[
\frac{1}{3\,\Omega}\,\left(m^2 - \frac{2\lambda}{3}\right)
\psi^2\,\nabla_{[\rho}\Omega\,g_{\mu] \nu}
\]
\[
+ \left(\frac{4}{3}\,s\,\psi^2 
+ \frac{1}{3}\,\psi\,\nabla_{\pi}\Omega\,\nabla^{\pi}\psi
\right)\nabla_{[\rho}\Omega\,g_{\mu] \nu}
+ \psi\,\nabla_{[\rho}\Omega\,\nabla_{\mu]}\psi\,\nabla_{\nu}\Omega
- \frac{\lambda}{3}\,\psi\,\nabla_{[\rho}\psi\,g_{\mu] \nu}
\]
\[
+ \Omega\,\left\{\nabla_{[\rho}\Omega\,\nabla_{\mu]}\psi\,\nabla_{\nu}\psi
+ \left(2\,s\,\psi + \nabla_{\pi}\Omega\,\nabla^{\pi}\psi
+ \frac{\kappa}{12}\Omega\,\psi\,T_*\right)\nabla_{[\rho}\psi\,g_{\mu] \nu}
\right.
\]
\[
\left.
+ \left(\frac{\kappa}{18}\,\psi^2\,T_* - \frac{1}{3}\,\nabla_{\pi}\psi\,\nabla^{\pi}\psi
\right)\nabla_{[\rho}\Omega\,g_{\mu] \nu}\right\}
\]
\[
+ \frac{2}{3}\,\Omega^{-3}\,V(\Omega\,\psi)\,\nabla_{[\rho}\Omega\,g_{\mu] \nu},
\]
and thus finally
\begin{equation}
\label{sing-field}
\hat{\nabla}_{[\rho} \hat{L}_{\mu] \nu} =  
\frac{\kappa}{2}\left[
 \frac{1}{3}\left(1 + \frac{\psi}{\Omega}\right)
\left(m^2- \frac{2\,\lambda}{3}\right)\psi\,\nabla_{[\rho}\psi\,g_{\mu]\,\nu}
+ \Omega\,K_{\rho \mu \nu}
\right],
\end{equation}
with the field
\[
K_{\rho \mu \nu} = 
\]
\[
+ 
\left(\frac{\kappa}{18}\,\psi^2\,T_* 
- \frac{2}{3}\,\nabla_{\pi}\psi\,\nabla^{\pi}\psi\right)\nabla_{[\rho}\Omega\,g_{\mu] \nu}
+ \left(\frac{1}{2}\,s\,\psi
+ \frac{2}{3}\,\nabla_{\pi}\Omega\,\nabla^{\pi}\psi
+ \frac{\kappa}{12}\Omega\,\psi\,T_*\right)\nabla_{[\rho}\psi\,g_{\mu] \nu}
\]
\[
- \psi\,\nabla_{[\rho}\Omega\,\nabla_{\mu]} \nabla_{\nu}\psi
+ \psi^2\,\nabla_{[\rho}\Omega\,L_{\mu] \nu}
+ \frac{1}{3}\,\psi^2\,\nabla^{\pi}\Omega\,L_{\pi[\rho}\,g_{\mu]\nu}
- \frac{1}{3}\,\psi\,\nabla^{\pi}\Omega\,\nabla_{\pi} \nabla_{[\rho}\psi\,g_{\mu] \nu}
\]
\[
- \frac{\kappa}{2}\psi^2\,\nabla_{[\rho}\Omega\,T^*_{\mu]\nu} 
-  \frac{\kappa}{6}\psi^2\,\nabla^{\pi}\Omega\,T^*_{\pi [\rho}\,g_{\mu] \nu} 
+ 2\,\nabla_{[\rho}\Omega\,\nabla_{\mu]}\psi\,\nabla_{\nu}\psi 
\]
\[
+ \Omega\left\{
- \nabla_{[\rho}\psi\,\nabla_{\mu]}\nabla_{\nu}\psi 
-\frac{1}{3}\,\nabla^{\pi}\psi\,\nabla_{\pi}\nabla_{[\rho}\psi\,g_{\mu]\nu} 
-\frac{\kappa}{36}\,\psi\,T_*\,\nabla_{[\rho}\psi\,g_{\mu]\nu}
\right.
\]
\[
\left. -\frac{\kappa}{2}\,\psi\,\nabla_{[\rho}\psi\,T^*_{\mu] \nu}
+  \frac{\kappa}{12}\,\psi\,\nabla^{\pi}\psi\,T^*_{\pi [\rho}\,g_{\mu] \nu}
+ \psi\,\nabla_{[\rho}\psi\,L_{\mu]\nu} 
- \frac{1}{6}\,\psi\,\nabla^{\pi}\psi\,L_{\pi [\rho}\,g_{\mu] \nu}
 \right\}
\]
\[
+ \frac{1}{3}\,
 \Omega^{-3}\,\psi\,V'(\Omega\,\psi)\,\nabla_{[\rho}\Omega\,g_{\mu] \nu}
+ \frac{1}{3}\,\Omega^{-2}\,V'(\Omega\psi)\,\nabla_{[\rho}\psi\,g_{\mu] \nu},
\]
which is regular as $\Omega \rightarrow 0$. From this our assertion follows immediately. $\Box$

\section{The constraints on a hypersurface $\{\Omega = \,0\}$.}
\label{constraints}

In \cite{friedrich:1986a}  it has been observed that the constraints induced by the conformal vacuum field equations with positive cosmological constant  simplify on a hypersurface  
 ${\cal J} = \{\Omega = \,0\}$. Cauchy data for the conformal field equations on such a hypersurface are referred to as 
{\it asymptotic initial data}.  In the following  it will be shown that also in the case of the Einstein-scalar field system satisfying $(*)$ 
the construction of asymptotic initial data is considerably simpler than the construction of standard Cauchy data for the Einstein-scalar field system (cf. \cite{choque-bruhat-isenberg-pollack:2007}).

To derive and analyse the constraints, the compact manifold ${\cal J}$ will be thought of as being  embedded as a Cauchy hypersurface into a smooth solution to the conformal field equations. It will be convenient to assume the
solution metric $g$ to be given in terms of Gauss coordinates based on 
${\cal J}$ so that  $\Omega > 0$ in the past  and $\Omega < 0$ in the future of ${\cal J}$ close to it. Then 
\[
g = - d\tau^2 + h_{ab}\,dx^a\,dx^b,
\]
with $x^0 \equiv \tau = 0$ and $\partial_{\tau}\Omega < 0$  on ${\cal J}$.
The Christoffel symbols are given by
\[
\Gamma_a\,^0\,_b[g] = \frac{1}{2}\,h_{ab,0} = \chi_{ab}, \quad 
\Gamma_0\,^0\,_b[g] = 0, \quad \Gamma_a\,^0\,_0[g] = 0, \quad \Gamma_0\,^c\,_0[g] = 0,
\]
\[
\Gamma_a\,^c\,_0[g] = \Gamma_0\,^c\,_a[g] = h^{ce}\,\chi_{eb}, \quad 
\quad \Gamma_a\,^c\,_b[g] = \Gamma_a\,^c\,_b[h].
\]
where $\chi_{ab}$ denotes the second fundamental form 
and $h^{ab}$ the inverse of the metric  $h_{ab}$  induced on  $\{\tau = const.\}$. With this notation we can state the following result.

\begin{proposition}
\label{constr-data-on-scri}
Assume $\lambda > 0$. On a smooth, orientable, compact $3$-manifold ${\cal J}$ let be given a smooth Riemannian metric $h_{ab}$ with covariant derivative operator $D_a$, and smooth scalar fields $\psi_0$,  $\psi_1$ so that 
\begin{equation}
\label{solvability-cond}
\int_S X^a\,\rho_a\,d\mu_h = 0 \quad \mbox{
for any conformal vector field $X$ admitted by $h$},
\end{equation}
where
\[
\rho_a = \frac{\kappa}{3}\,\Sigma\,( \psi_0\,D_{a}\psi_1 - 2\,\psi_1 \,D_{a}\psi_0) \quad \mbox{with} \quad
\Sigma =  - \sqrt{\frac{\lambda}{3}}.
\]
Furthermore let $w_{ab}$ be a smooth, symmetric, trace free solution to the equation
\begin{equation}
\label{div(w)-equ}
D^aw_{ab} = \rho_b.
\end{equation}
Initial data (i.e. solutions to the constraints) for the conformal Einstein-scalar field equations  with cosmological constant $\lambda$ and mass $m > 0$ satisfying the condition $(*)$ 
are then derived  in a suitable conformal gauge from these data on ${\cal J}$ as follows. 

\noindent
$-$ The fields $\psi_0$,  $\psi_1$ constitute the Cauchy data $\psi$ and $\nabla_0\psi$ for the scalar field $\psi$ on ${\cal J}$.\\
$-$ The inner metric and the second fundamental form on ${\cal J}$ are given by
\[
h_{ab} \quad \mbox{and} \quad \chi_{ab} = 0,
\]
$-$ The conformal factor, its time derivative, and the function $s$  are given by
\[
\Omega = \,0, \quad \nabla_0\Omega = \, \Sigma, \quad s = 0.
\]
$-$ The Schouten tensor of $g$  is given by
\[
L_{ab}[g] = L_{ab}[h], \quad L_{0a}[g] = L_{a0}[g] = 0, \quad 
L_{00}[g] = \frac{1}{4}\,R[h] - \frac{1}{6}\,R[g],
\]
where $R[g]$ is to be considered as a smooth conformal gauge source function which can be given arbitrarily.\\
$-$ The rescaled conformal Weyl tensor $W^{\mu}\,_{\nu \lambda \rho}$ is specified in terms of its 
electric part with respect to ${\cal J}$, which is given by $w_{ab}$, and its magnetic part which is given by
\[
w^*_{cd} = - \frac{1}{\Sigma}\,D_{a}\,L_{b c}[h]\,\epsilon_d\,^{ab}. 
\]
\end{proposition}

\vspace{.2cm}

\noindent
{\bf Proof:}
Regularity of $T_*$ and the relation $\hat{T} = \Omega^2\,T_*$
imply that 
\[
\hat{T} = 0, \quad \hat{\nabla}_{\mu}\,\hat{T} = 0 \quad \mbox{on} \quad {\cal J}.
\]
The restriction of  (\ref{mat-B-Ric-scal-transf}) to $ {\cal J}$ thus gives
\begin{equation}
\label{dOmega-on-scri}
\nabla_0\Omega =  \Sigma =  - \sqrt{\frac{\lambda}{3}}.
\end{equation}
The restriction of (\ref{mat-B-Ric-ten-transf}) with $\mu = a$, $\nu = 0$ is satisfied 
because of the values of the Christoffel symbols and because 
$\nabla_0\Omega$ is  constant on $ {\cal J}$ while it  gives with $\mu = a$, $\nu = b$
 the relation
\[
0 = \nabla_{a}\,\nabla_{b}\Omega - s\,g_{ab} = - \chi_{ab}\,\Sigma - s\,h_{ab}, 
\]
which implies that the trace-free part of $\chi_{ab}$ vanishes and its trace is given by
$\chi = - \Sigma^{-1}\,s$.
So far we did not make use of the conformal gauge freedom. It allows us to perform 
arbitrary rescalings with positive conformal factors $\theta$ and can be removed by prescribing the Ricci scalar $R[g]$ as given  function near ${\cal J}$ and by prescribing $\theta$ and its time derivative 
on  ${\cal J}$. Leaving the freedom to choose $\theta$ on ${\cal J}$ untouched,  its time derivative can be chosen there to achieve 
\[
\chi = 0 \quad \mbox{whence} \quad s = 0 \quad \mbox{and} \quad \chi_{ab} = 0
\quad \mbox{on} \quad {\cal J}.
\]

Observing now that $C^{\mu}\,_{\nu \rho \lambda} = \Omega\,W^{\mu}\,_{\nu \rho \lambda} = 0$
on $ {\cal J}$ the last relation implies with Gauss' equation that
\[
\frac{1}{2}\,R_{ab}[g] - \frac{1}{12}\,R[g]\,h_{ab} = L_{ab}[g] = L_{ab}[h] \equiv R_{ab}[h] - \frac{1}{4}\,R[h]\,h_{ab}
\quad \mbox{on} \quad {\cal J}, 
\]
and thus by contraction
\[
R[g] + 2\,R_{00}[g] = R[h],
\]
whence
\[
L_{00}[g] = \frac{1}{2}\,R_{00}[g] - \frac{1}{12}\,R[g]\,g_{00} = \frac{1}{4}\,R[h] - \frac{1}{6}\,R[g].
\]
It holds
\[
T^*_{\mu\nu} 
= \psi^2(\nabla_{\mu}\Omega\,\nabla_{\nu}\Omega 
- \frac{1}{4}\,\nabla_{\pi}\Omega\,\nabla^{\pi}\Omega\,g_{\mu\nu})
\quad \mbox{on} \quad {\cal J}.
\]
The constraint induced by (\ref{mat-s-equ}) with  $\mu = a$ 
reduces thus to
\[
L_{0a}[g] = L_{a0}[g] = 0.
\]
It follows that the initial data for $L_{\mu\nu}[g]$ can be expressed completely in terms of 
$L_{ab}[h]$ and the gauge dependent quantity $R[g]$ which can be prescribed arbitrarily near ${\cal J}$.

Because $\hat{\nabla}_{[\nu}\,\hat{L}_{\lambda] \rho} = 0$ on $ {\cal J}$ by 
 (\ref{scal-field-reg-cond}) and
(\ref{sing-field}), the constraints induced by (\ref{mat-L-equ})
are given by 
\[
\nabla_{a}\,L_{b \rho} 
- \nabla_{b}\,L_{a \rho} = 
 \Sigma\,\,W_{\rho 0 a b}.
\]
The relation with $\rho = 0$ is satisfied by the fields given above
and implies no condition. 
The remaining relation can be written in the form
\[
D_{a}\,L_{b c}[h] 
- D_{b}\,L_{a c}[h] = 
 \Sigma\,\,w^*_{cd}\,\epsilon^d\,_{a b},
\]
where $D_a$ denotes the $h$-covariant derivative operator 
and $w^*_{ab} \equiv - \frac{1}{2}\,W_{a0cd}\,\epsilon_b\,^{cd}$ the
${\cal J}$- magnetic part of $W^{\mu}\,_{\nu \rho\lambda}$.
It is saying that the magnetic part of the rescaled conformal Weyl tensor is given on ${\cal J}$ by the Cotton tensor defined by $h_{ab}$.

One of the constraints implied by (\ref{mat-W-equ}) is obtained   by restricting 
\[
\nabla_{\mu}\,W^{\mu}\,_{0 a b} 
= 2\,\Omega^{-1}\,\hat{\nabla}_{[a}\,\hat{L}_{b] 0} = \kappa\,K_{ab0},
\]
to $ {\cal J}$. With the results above it follows that the restriction of $K_{ab0}$ to $ {\cal J}$ vanishes and the constraint reduces to 
\[
D^a\,w^*_{ab} = 0.
\]
This is just  the 
differential identity satisfied by the Cotton tensor and gives no new condition.  
The only remaining constraint is given by the restriction to ${\cal J}$ of 
\[
\nabla_{\mu}\,W^{\mu}\,_{0 a 0} 
= 2\,\Omega^{-1}\,\hat{\nabla}_{[a}\,\hat{L}_{0] 0} = \kappa\,K_{a00},
\]
which can be written with the results above 
\[
D^b\,w_{ba} = \rho_a \quad \mbox{with} \quad 
\rho_a = \frac{\kappa}{3}\,\Sigma\,( 
\psi_0\,\nabla_{a} \nabla_{0}\psi
- 2\,\nabla_{0}\psi \,\nabla_{a}\psi).
\]
Because
\[
\nabla_{a} \nabla_{0}\psi = \partial_{a} \nabla_{0}\psi - \chi_{a}\,^c\, D_{c}\psi
= \partial_{a} \partial_{0}\psi,
\]
the field $\rho_a$ can be expressed completely in terms of $\kappa$, $\Sigma$, and the data
$\psi_0$ and $\nabla_0\psi$ for the scalar field on  $ {\cal J}$.
For any smooth vector field $X^a$ on ${\cal J}$ holds
\[
\int_S X^a\,\rho_a\,d\mu_h = \int_S X^a\,D^b\,w_{ab}\,d\mu_h
= \int_S w_{ab}\,(D^{(b}X^{a)} - \frac{1}{3}\,D_cX^c\,h^{ab})\,d\mu_h.
\]
because $w_{ab}$ is trace free. It follows that the data $h$, $\psi$, and $\nabla_0\psi$ must  be given such that (\ref{solvability-cond}) holds true. If this condition is satisfied the well known properties of the operator 
$\mathbb{L}_h = div \circ {\cal L}_h$, where the divergence of a covariant symmetric trace free tensor field $w_{ab}$  is the 1-form $(div \,w)_b = - D^a w_{ab}$ and
the conformal Killing operator ${\cal L}_h$ acts on a 1-form $X_a$ by 
$ ({\cal L}_h\;X)_{ab} = D_{(b}X_{a)} - \frac{1}{3}\,D_cX^c\,h_{ab}$, then imply that the equation
$\mathbb{L}_h \,X = \rho$ is solvable and a solution to (\ref{div(w)-equ}) is provided by 
the tensor $ ({\cal L}_h\;X)_{ab} $ calculated from $X$.\\
$\Box$

\vspace{.2cm}

\noindent
{\bf Remarks:}

All possible smooth asymptotic initial data for which $\Omega$ is decreasing near ${\cal J}$  are obtained by the procedure described above.

The relation  (\ref{dOmega-on-scri})     (already observed in \cite{penrose:1965}) shows that the set 
$\{\Omega = 0\}$ is necessarily a space-like hypersurface.

The  initial data for the wave equation which needs to be derived for $\nabla_{\mu}\psi$ from
(\ref{conf-K-G-equ})  are given by
$\nabla_a\psi = \partial_a\psi_0$, $\nabla_0\psi = \psi_1$, 
$\nabla_0\nabla_a\psi = \nabla_a\nabla_0\psi = \partial_a\psi_1$, and the datum $\nabla_0\nabla_0\psi$, which can be read off from (\ref{conf-K-G-equ}), is found to be 
\[
\nabla_0\nabla_0\psi = h^{ab}\,D_aD_b\psi_0 - \frac{1}{6}\,R[g]\,\psi_0 
+ \left(\frac{\kappa\,\lambda}{6}
- 4\,U(0)\right)\psi^3_0.
\]

Given an asymptotic initial data set as above and a space-time gauge in which the conformal field equations imply hyperbolic evolution equations, the latter allow us to determine a past Cauchy development of the data which provides, where $\Omega > 0$, a 
unique maximal, globally hyperbolic, future asymptotically simple solution $\hat{g} = \Omega^{-2}\,g$, $\phi = \Omega\,\psi$ of the system  (\ref{I-Einst-equ}), (\ref{I-K-G-equ}), (\ref{I-K-G-em-tensor}) with (\ref{scal-field-reg-cond}). 

There exists, however, also a unique maximal, globally hyperbolic future Cauchy development of the data on which $\Omega < 0$. The conformal field equations are left invariant by the transition 
under which $\Omega \rightarrow - \Omega$, 
$W^{\mu}\,_{\nu \lambda \rho} \rightarrow - W^{\mu}\,_{\nu \lambda \rho}$,
$\psi \rightarrow - \psi$ while all other unknowns remain unchanged.
The fields $\hat{g} = \Omega^{-2}\,g$, $\phi = \Omega\,\psi$ obtained from the future development thus define again a solution to  (\ref{I-Einst-equ}), (\ref{I-K-G-equ}), (\ref{I-K-G-em-tensor}) with (\ref{scal-field-reg-cond}), which now is asymptotically simple in the past.

Because the metric $h_{ab}$ is not subject to an analogue of the Hamiltonian constraint, 
neither the topology of ${\cal J}$ nor the conformal structure defined by $h_{ab}$ is  restricted in any way.
It may appear then that the procedure offers too much freedom. This is not the case. As discussed in the proof, there remains the freedom to perform on ${\cal J}$ transitions of the form
\[
\Omega \rightarrow \Omega^* = \theta\,\Omega, \quad
g \rightarrow g^* = \theta^2\,g,
\]
with positive conformal factors $\theta$. The effect of these rescalings on the free data is
\[
h_{ab} \rightarrow h^*_{ab} = \theta^2\,h_{ab}, \quad 
\psi_0 \rightarrow \psi^*_0 =\theta^{-1}\,\psi_0, 
\]
\[
\psi_1 \rightarrow \psi^*_1 = \theta^{-2}\,\psi_1, \quad
w_{ab} \rightarrow w^*_{ab} = \theta^{-1}\,w_{ab}.
\]
These rescalings change the conformal representation of the solution to the conformal field equations but leave the  associated `physical solution' unchanged.

To assess the freedom to prescribe data one should observe that
(\ref{div(w)-equ}) leaves the freedom to add 
to a given solution of the inhomogeneous equation an arbitrary solution of the homogeneous equation $D^a\,w_{ab} = 0$.
In spite of the condition $\chi_{ab}  = 0$ on ${\cal J}$, which reflects the particular nature of the conformal boundary, the procedure described above thus admits essentially the same freedom to prescribe data as the standard Cauchy problem for the system 
(\ref{I-Einst-equ}), (\ref{I-K-G-equ}), (\ref{I-K-G-em-tensor}) with (\ref{scal-field-reg-cond}). This conclusion is supported by the following observation.

Denote by $S$  a smooth $3$-manifold diffeomorphic to ${\cal J}$, 
by ${\cal D}_S$ the set of smooth Cauchy data on $S$ for the system  (\ref{I-Einst-equ}), (\ref{I-K-G-equ}), (\ref{I-K-G-em-tensor}) satisfying $(*)$, 
and by ${\cal A}_S$ the subset of the Cauchy data $d_S \in {\cal D}_S$ for which the corresponding solution space-times develops so as to admit a smooth conformal boundary ${\cal J}^+ \sim S \sim {\cal J}$ in the future. $S$ can be thought of as being embedded as a Cauchy hypersurface into this solution so that the data induced by the solution are isometric to the data  $d_S$.
The solution induces on ${\cal J}^+$ an asymptotic initial data set 
$d_S^*$ for the conformal field equations. The past development of the data $d_S^*$ with the conformal field equations is conformal to the solution developed from $d_S$. It induces on $S$ data
$d'_S$ for the conformal field equations. As discussed above, the future evolution of these data by the conformal field equations extends smoothly beyond ${\cal J}^+$ into a domain which can be foliated by Cauchy hypersurfaces on which 
$\Omega < 0$.

With data $\bar{d}_S$ on $S$ which are obtained by a small 
(non-linear) perturbation of $d_S$ we can associate data $\bar{d}'_S$ for the conformal field equations which represent a small perturbation of $d'_S$ so that in terms of suitable Sobolev topologies
$\bar{d}'_S \rightarrow d'_S$ as $\bar{d}_S \rightarrow d_S$ and vice versa.
Cauchy stability for the conformal field equation implies that data $\bar{d}'_S$ 
which are close enough to $d'_S$ will then also develop into a domain  foliated by Cauchy hypersurfaces on which $\Omega' < 0$ and the analogue of 
(\ref{dOmega-on-scri}) for $\Omega' $, which is a consequence of the field equations where $\Omega' = 0$, ensures that the set $\{\Omega' = 0\}$ is a smooth hypersurface
diffeomorphic to $S$. Solutions arising from data close enough to $d_S$ will thus be asymptotically simple in the future and belong to ${\cal A}_S$. {\it The set 
${\cal A}_S$ is thus open in the set ${\cal D}_S$} if the latter is endowed with a suitable Sobolev topology. By the results of \cite{friedrich:1991} this statement can be generalized to include perturbations involving 
fields with trace free energy momentum tensor which  satisfy conformally covariant field equations.

\section{Spatially homogeneous solutions.}
\label{hom-sols}

In the following we study solutions to the equation considered in Theorem \ref{main-result}
for which the conformal factor $\Omega$ and the metric $g$ are defined on $\mathbb{R} \times S$ with $S = \mathbb{S}^3, \, \mathbb{T}^3$ or $\mathbb{H}^3_*$ (a factor space of hyperbolic $3$-space) and take the form
\[
\Omega = \Omega(\tau), \quad \quad g = - d\tau^2 + l^2\,k,
\]
with a function $l = l(\tau)$ and a $3$-metric  $k =  k_{ab}\,dx^a\,dx^b = k_{\epsilon}$  
of constant curvature $R_{abcd}[k] = 2\,\epsilon\,k_{a[c}\,k_{b]d} $, where $\epsilon = 1, 0, -1$ respectively.
We write also $\tau = x^0$ and assume 
for simplicity 
\[
V = 0, \quad \quad \kappa = 1,  \quad \quad \lambda = 3,  
\]
and thus $m^2 = 2$ to take care of 
(\ref{scal-field-reg-cond}). The non-vanishing Christoffel symbols  and the second fundamental form $\chi_{ab}$ of the slices $\{\tau = const.\}$ are then given by
\[
\chi_{ab} = \Gamma_a\,^0\,_b[g] = l\,l'\,k_{ab},  \quad
\Gamma_0\,^a\,_c[g] = \Gamma_c\,^a\,_0[g] = \frac{1}{l}\,l'\,k^a\,_c, \quad
\Gamma_b\,^a\,_c[g] =  \Gamma_b\,^a\,_c[k], 
\]
where $' = \frac{d}{d\tau}$. The Ricci scalar and the Ricci tensor are given by
\[
R[g] = \frac{6}{l^2}\,(\epsilon + l\,l'' + (l')^2), \quad \quad 
R_{00}[g] = - 3\,\frac{l''}{l}, \quad \quad R_{a0}[g] = R_{0a}[g] = 0,
\]
\[
R_{ab}[g] = \{2\,\epsilon + l\,l'' + 2\,(l')^2\}\,k_{ab},
\]
and the Schouten tensor by
\[
L_{00}[g] = \frac{1}{2\,l^2}\,(\epsilon - 2\,l\,l'' + (l')^2), \quad 
L_{a0}[g] = L_{0a}[g] = 0, \quad 
L_{ab}[g] = \frac{1}{2}\,(\epsilon + (l')^2)\,k_{ab}.
\]
Because the line element above is conformally flat it follows that 
$W^{\mu}\,_{\nu \lambda \rho} = 0$, which leads to a considerable simplification.

It will be assumed in the following that the conformal time coordinate $\tau$ vanishes on a set  
$\{\Omega = 0\}$ and the metric $g$ satisfies the conformal  gauge condition
$R[g] = 6\,\epsilon$.
Fixing $R$ still leaves some freedom to perform  rescalings. This can be used to restrict the metric and the second fundamental form on $\{\Omega = 0\}$ so that $l =1$ and $l' =0$ there. With these requirements and the expression for the Ricci scalar above follows that 
\[
l\,l'' + (l')^2 + \epsilon\,(1 - l^2) = 0, \quad l(0) = 1, \quad l'(0) = 0,
\]
which implies that $l = 1$. Where $\Omega > 0$ the physical fields can then  be given in the form
\begin{equation}
\label{hom-g-hat}
\hat{g} = \Omega^{-2}\,g = - dt^2 + f^2\,d\omega^2, \quad \quad \phi = \Omega\,\psi,
\end{equation}
with
\begin{equation}
\label{coord-conf-transf}
f(t) = \frac{1}{\Omega(\tau(t))}, \quad \quad 
\frac{dt}{d\tau} = \frac{1}{\Omega(\tau)},
\end{equation}
so that the information on the geometry is completely encoded in the conformal factor.

Because $g$ is conformally flat we have by (\ref{contr-Bianchi}) and (\ref{hat-contr-Bianchi})
$\nabla_{[\nu}\,L_{\lambda] \rho} = 0$ and 
$\hat{\nabla}_{[\nu}\,\hat{L}_{\lambda] \rho} = 0$ so that equations 
(\ref{mat-L-equ}), (\ref{mat-W-equ}) are trivially satisfied and we are left with the
equations
\begin{equation}
\label{C-mat-B-Ric-scal-transf}
2\,\Omega\,s - \nabla_{\rho}\Omega\,\nabla^{\rho}\Omega = 
1 - \frac{1}{12}\,\Omega^2\,T_*,
\end{equation}
\begin{equation}
\label{C-mat-B-Ric-ten-transf}
\nabla_{\mu}\,\nabla_{\nu}\Omega = - \,\Omega\,L_{\mu\nu} + s\,g_{\mu\nu}
+ \frac{1}{2}\,\Omega\,T^*_{\mu\nu},
\end{equation}
\begin{equation}
\label{C- mat-s-equ}
\nabla_{\mu}\,s = - \,\nabla^{\rho}\Omega\,L_{\rho\mu} 
+ \frac{1}{2}\,\nabla^{\rho}\Omega\,T^*_{\rho \mu}
- \frac{1}{12}\,\nabla_{\mu}\Omega\,T_*
- \frac{1}{24}\,\Omega\,\nabla_{\mu}T_*,
\end{equation}
\begin{equation}
\label{C-conf-K-G-equ}
\Box_g \psi - \frac{1}{6}\,R\,\psi 
= \frac{1}{6}\,T_* \psi,
\end{equation}
where
\begin{equation}
\label{C-trace-free-part-e-m-tensor}
T^*_{\mu\nu} 
= \nabla_{\mu}(\Omega\psi)\,\nabla_{\nu}(\Omega\,\psi)
- \frac{1}{4}\,\nabla_{\rho}(\Omega\psi)\,\nabla^{\rho}(\Omega\psi)\,\,g_{\mu\nu},
\end{equation} 
 \begin{equation}
\label{C-T_*-behaviour}
T_* = 
- \nabla_{\rho}(\Omega\,\psi)\,\nabla^{\rho}(\Omega\,\psi) - 4\,\psi^2.
\end{equation}
The assumed symmetry implies
\begin{equation}
\label{RC-final-T_*-behaviour}
T_* = ((\Omega\,\psi)')^2 - 4\,\psi^2,
\end{equation}
\begin{equation}
\label{RC-trace-free-part-e-m-tensor}
T^*_{00} = \frac{3}{4}\,((\Omega\,\psi)')^2,
\quad \quad T^*_{a0}  = T^*_{0a}  = 0, \quad\quad 
T^*_{ab} 
= \frac{1}{4}\,((\Omega\psi)')^2\,k_{ab},
\end{equation} 
and the equations reduce to
\begin{equation}
\label{RC-mat-B-Ric-scal-transf}
2\,\Omega\,s + (\Omega')^2 = 
1 - \frac{1}{12}\,\Omega^2\,T_*,
\end{equation}
\begin{equation}
\label{00-RC-mat-B-Ric-ten-transf}
\Omega'' = - \frac{1}{2}\,\epsilon\,\Omega- s
+ \frac{3}{8}\,\Omega\,((\Omega\,\psi)')^2,
\end{equation}
\begin{equation}
\label{ab-RC-mat-B-Ric-ten-transf}
0 = - \frac{1}{2}\,\epsilon\,\Omega + s
+ \frac{1}{8}\,\Omega\,((\Omega\psi)')^2,
\end{equation}
\begin{equation}
\label{RC- mat-s-equ}
s' = \frac{1}{2}\,\epsilon\,\Omega'
- \frac{3}{8}\,\Omega' \,((\Omega\,\psi)')^2
- \frac{1}{12}\,\Omega'\,T_*
- \frac{1}{24}\,\Omega\,T_*',
\end{equation}
 \begin{equation}
\label{RC-conf-K-G-equ}
- \psi'' - \epsilon\,\psi 
= \frac{1}{6}\,T_* \psi.
\end{equation}
Obviously there is some redundancy in this system.
Solving for the last term on the right  hand side of (\ref{ab-RC-mat-B-Ric-ten-transf}) inserting the result in the last term on the right hand side of 
 (\ref{00-RC-mat-B-Ric-ten-transf})  gives
$s = \frac{1}{4}( - \Omega'' + \epsilon\,\Omega)$,
which is just (\ref{s-def}). Conversely, this expression for $s$ implies with 
 (\ref{00-RC-mat-B-Ric-ten-transf}) the relation  (\ref{ab-RC-mat-B-Ric-ten-transf}).
Solving (\ref{ab-RC-mat-B-Ric-ten-transf})  instead for $s$ and using this to replace $s$ in 
 (\ref{00-RC-mat-B-Ric-ten-transf}) gives 
\begin{equation}
\label{fin-00-RC-mat-B-Ric-ten-transf}
\Omega''  + \left(\epsilon - \frac{1}{2}\,((\Omega\,\psi)')^2\right) \Omega = 0.
\end{equation}
Using  (\ref{RC-final-T_*-behaviour}) in (\ref{RC-conf-K-G-equ}) gives
\begin{equation}
\label{fin-B-conf-K-G-equ}
\psi'' + \left(\epsilon + \frac{1}{6}\,((\Omega\,\psi)')^2 - \frac{2}{3}\,\,\psi^2\right)\psi = 0.
\end{equation}
Inserting $s$ from  (\ref{ab-RC-mat-B-Ric-ten-transf}) in (\ref{RC-mat-B-Ric-scal-transf}) gives  
\begin{equation}
\label{fin-RC-mat-B-Ric-scal-transf}
(\Omega')^2 = 1 + \Omega^2\left(\frac{1}{3}\,\psi^2 +
\frac{1}{6}\,((\Omega\,\psi)')^2- \epsilon \right).
\end{equation}
The first two equations above  provide a closed evolution system for $\Omega$ and $\psi$ while the third equation should be read as a constraint. It will be  satisfied if it holds for one value of  $\tau$ and the first two equations are satisfied (not an immediate calculation).
 It can be shown that (\ref{RC- mat-s-equ}) follows if the other equations are satisfied.
By using (\ref{coord-conf-transf}), equations (\ref{fin-00-RC-mat-B-Ric-ten-transf}), (\ref{fin-B-conf-K-G-equ}),  (\ref{fin-RC-mat-B-Ric-scal-transf}) can be derived where $\Omega \neq 0$ directly from the equations implied by (\ref{Einst-equ}), (\ref{K-G-equ}), (\ref{K-G-em-tensor})
for the functions in (\ref{hom-g-hat}).

If data are prescribed on $\{\Omega = 0\} = \{\tau = 0\}$ and 
if is assumed that $\Omega$ is positive short before that crossover surface, the constraint (\ref{fin-RC-mat-B-Ric-scal-transf}) shows that the  equations above  must be solved with initial conditions
\[
\Omega(0) = 0, \,\,\, \Omega'(0) = - 1 
\quad \mbox{and free data} \quad \psi(0), \,\,\, \psi'(0).
\]
The vacuum solutions, obtained by setting  $\psi(0) = 0$, $\psi'(0) = 0$, satisfy  $\psi = 0$ and
$\Omega = \Omega_{\epsilon}$ with
\[
\Omega_1 = - \sin \tau,  \,\,\, - \pi < \tau < 0, \quad  
\Omega_0 = - \tau, \,\,\, \tau < 0, \quad 
\Omega_{-1} = - \sinh \tau, \,\,\, \tau < 0.
\]
Observing (\ref{coord-conf-transf}), the corresponding physical solutions 
$\hat{g}_{\epsilon} = \Omega^{-2}_{\epsilon}\,g_{\epsilon}$ are then given by the de Sitter solution
\[
\hat{g}_{1} = - dt^2 + \cosh^2t\,\,k_1, \quad t \in \mathbb{R},
\]
which expands in both time directions, and by
\[
\hat{g}_{0} = - dt^2 + e^{2t}\,k_0, \quad t \in \mathbb{R}, \quad \quad \quad 
\hat{g}_{-1} = - dt^2 + \sinh^2t\,\,k_{-1}, \quad t > 0.
\]

When $\psi \not\equiv 0$ it follows by (\ref{fin-00-RC-mat-B-Ric-ten-transf}) 
 in the cases $\epsilon \le 0$  that the functions
$ \Omega_{\epsilon}(\tau)$ grow even faster backwards in time and most likely diverge for a finite value 
$\tau_* < 0$ of the parameter $\tau$. The precise behaviour of the solutions near that point, 
or, in other words, the precise expansion behaviour of the associated physical solutions 
$\hat{g}_{\epsilon} = \Omega^{-2}\,g_{\epsilon}$, $\epsilon = 0, -1$, at the big bang indicated by that point,
depends on the initial data for $\psi$ and requires a detailed analysis which cannot be given here.

{\it In the following we consider the case $\epsilon = 1$} in somewhat more detail {\it and drop the index $\epsilon$ everywhere}.
The structure of the system (\ref{fin-00-RC-mat-B-Ric-ten-transf}), 
(\ref{fin-B-conf-K-G-equ}), (\ref{fin-RC-mat-B-Ric-scal-transf}) implies
 that for data
$\psi$ and $\psi'$ at $\tau = 0$ which are sufficiently small  the solutions $\Omega(\tau)$ will also be oscillatory and may stay close to the de Sitter solution for a long conformal time $\tau$.
It is an interesting problem to characterize  the initial data $\psi(0)$ and $\psi'(0)$ for which the solutions to the system system (\ref{fin-00-RC-mat-B-Ric-ten-transf}), 
(\ref{fin-B-conf-K-G-equ}), (\ref{fin-RC-mat-B-Ric-scal-transf}) exist for all conformal times $\tau$.
That there do exist non-trivial data with this property other than oscillatory  solutions is shown by the solution $\Omega_*$, $ \psi_*$ to
(\ref{fin-00-RC-mat-B-Ric-ten-transf}), 
(\ref{fin-B-conf-K-G-equ}), (\ref{fin-RC-mat-B-Ric-scal-transf})
which is given by 
\begin{equation}
\label{conf-is-co-sing}
\Omega_* = - \tau, \quad \psi_* = \sqrt{2}.
\end{equation}
The physical fields corresponding to the restriction of this solution to the domain 
$- \infty < \tau < 0$ are given in terms of the coordinate $ t = - \log (- \tau)$ by
\[
\tilde{g} = - dt^2 + e^{2t}\,d\omega^2,
\quad \phi = \sqrt{2}\,e^{-t},
\]
and are thus quite different from those in the de Sitter case.
As $t \rightarrow - \infty$ the matter field diverges
while it decays and the metric shows a  de Sitter type expansion behaviour as $t \rightarrow \infty$.

Because $\Omega_*$ does not approach zero a second time but $\Omega_* \rightarrow \infty$
as $\tau \rightarrow - \infty$, it is not possible to obtain a global stability result by the argument used before, but the situation may be of considerable interest in our context.
Consider the backward evolution of initial data of the form
\[
\Omega(0) = 0, \quad \Omega(0) = - 1, \quad \psi(0) = \sqrt{2} - \delta, \quad
\psi'(0) = - \bar{\delta}, \quad 0 \le \delta < \sqrt{2}, \quad 0 \le \bar{\delta}. 
\]
Numerical calculations with data so that $0 \le \delta, \,\bar{\delta} << 1$
show that there exist solutions $\Omega$ which stay close to $\Omega_*$  for $\tau < 0$ and $|\tau|$ small enough
and which are monotonically increasing with $\Omega < - \tau$. 
After assuming a maximum value $\Omega_m$ at some $\tau_m < 0$, the solutions are decreasing until they vanish at some point $\tau_z < \tau_m$. The corresponding physical solutions on $] \tau_z, 0[$  can be thought of as arising from initial data on the `crossover surface' $\{\tau = \tau_z\}$, of developing  a `waist' of volume $\Omega_m^{-3}\,Vol(\mathbb{S}^3)$ at $\tau_m$,
and approaching the next crossover surface  at $\{\tau = 0\}$. Again these solutions would have the stability property pointed out above. First calculations show that by suitable choices of the data the value of 
$|\tau_m|$, the maximum value  $\Omega_m$, and the value of $|\tau_z|$ can be made to increase. This raises an interesting question, which is related to the second type of problems addressed in the introduction: 

\vspace{.2cm}

\noindent
{\it Do there exist solutions of this type which approximate for  given $z < 0$ the solution $\Omega_*$  on the  interval $[z, 0]$ arbitraryly well ?}

\vspace{.2cm}

\noindent 
A positive answer would show the existence of solutions which still have the stability property but whose waist would be arbitrarily narrow. The restriction of such solutions to the range $]\tau_m, 0[$ would, from the point of view of observational data, hardly be distinguishable from solutions which start with a big bang and then expand exponentially. Again, no attempt is made here to analyze this question.

\vspace{.5cm}

\noindent
{\it Acknowledgements}: I would like to thank Piotr Bizon and David Garfinkle  for discussions, and the Clay Mathematics Institute for financial support.
This material is partly based upon work supported by the National Science Foundation under Grant No. 0932078 000, while the author was in residence at the Mathematical Sciences Research Institute in Berkeley, California, during the year  2013.

}

\end{document}